\documentclass[11pt]{article}

\usepackage{amsmath}
\usepackage{graphicx}
\usepackage{enumerate}
\usepackage{natbib}
\usepackage{typearea}
\usepackage[hyphens]{url}
\usepackage{hyperref}

\hypersetup{
           breaklinks=true,   
           hidelinks,  
           pdfusetitle=true,  
        }

\usepackage{amsfonts, amsmath, amssymb, amsthm, dsfont}
\usepackage{amsthm}
\usepackage{natbib}
\usepackage{booktabs}
\usepackage{multirow}

\usepackage{bbm,bm}

\def\bfz{\mathbf{z}}

\def\Var{\textup{Var}}

\def\E{\mathsf{E}\,}

\def\inL2{\,{\buildrel L_2 \over \rightarrow}\,}

\newcommand{\bra}[1]{\left[#1\right]}
\newcommand{\cur}[1]{\left\{#1\right\}}
\newcommand{\pa}[1]{\left(#1\right)}
\newcommand{\abs}[1]{\left|#1\right|}

\title{Variational inference for max-stable processes}
\author{
Patrik Andersson\thanks{Uppsala University ({patrik.andersson@statistics.uu.se})}
\and  
Alexander Engberg\thanks{Uppsala University ({alexander.engberg@statistics.uu.se})}
} 

\date{}

\begin{document}

\maketitle

\begin{abstract}
Max-stable processes provide natural models for the modelling of spatial extreme values observed at a set of spatial sites. Full likelihood inference for max-stable data is, however, complicated by the form of the likelihood function as it contains a sum over all partitions of sites. As such, the number of terms to sum over grows rapidly with the number of sites and quickly becomes prohibitively burdensome to compute.

We propose a variational inference approach to full likelihood inference that circumvents the problematic sum. To achieve this, we first posit a parametric family of partition distributions from which partitions can be sampled. Second, we optimise the parameters of the family in conjunction with the max-stable model to find the partition distribution best supported by the data, and to estimate the max-stable model parameters.

In a simulation study we show that our method enables full likelihood inference in higher dimensions than previous methods, and is readily applicable to data sets with a large number of observations. Furthermore, our method can easily be extended to a Bayesian setting. Code is available at \url{https://github.com/LPAndersson/MaxStableVI.jl}.


\end{abstract}

\pagebreak

\section{Introduction}\label{sec1}

Max-stable processes are the only non-degenerate limits of rescaled component-wise maxima from independent and identically distributed stochastic processes \citep[ch 9]{deHaan2006}. Consequently, they constitute suitable models for spatial modelling of extreme events, such as high temperatures \citep{Davison2012a}, extreme levels of air pollution \citep{Vettori2019}, and heavy rainfall \citep{Huser2014}. Typically, spatial extreme values are observed at a number $D$ of spatial sites $ \{s_1,\dots,s_D\} $, where $ s_i\in \mathcal S \subset \mathbb R^2 $. From each site the component-wise maxima from blocks of data from some stochastic process $ X $, are recorded. That is, we observe
\begin{equation}\label{eq:blockmax}
	\max_{1\leq i\leq m}\{ X_i(s): s \in \mathcal S \},
\end{equation}
where $ m $ is the block size, and denote this block-maxima sample $ (x_{m,1},\dots,x_{m,D}) $.

Likelihood-based methods are often used to fit max-stable processes to data, due to their favourable large sample properties. Full likelihood inference is, however, complicated by the form of the likelihood which contains a sum of the $ D $th Bell number of terms. For $ D = 10 $ this amounts to more than $ 10^5 $ terms. Each term in the sum corresponds to a partition, $ \pi $, of the set $ \{1, \dots, D\} $ which specifies whether or not maxima observed at different sites occurred simultaneously and thus were caused by the same extremal event. Computing the full likelihood is computationally prohibitive already in moderate dimensions (about $ D $ between 5 and 10) \citep{Castruccio2016, Huser2019} whereby alternative methods are needed for full likelihood inference. 

Various attempts to circumvent the problematic sum have been proposed. \cite{Padoan2010} suggest using a composite likelihood approach in which pairwise likelihoods are fitted to data from pairs of sites which reduces model fitting to dimension $ D=2 $ while maintaining consistency, although at an efficiency loss. This was extended to dimensions higher than $ 2 $ \citep{Genton2011, Huser2013, Sang2014, Castruccio2016}, however full efficiency was not achieved.
Furthermore, composite likelihoods make it more difficult to assess the uncertainty and to adapt the models to a Bayesian setting \citep{Varin2011}. 
A different approach was proposed by \cite{Stephenson2005} who showed that by viewing $ \pi $ as a random variable, one can use the joint likelihood of the data and $ \pi $. This reduces the problematic sum to a single term. They further suggest using the empirical partition $ \hat\pi_m $, i.e.\ the partition implied by the occurrence times of the block maxima, as an observation of $ \pi $. \cite{Wadsworth2015}, however, showed that fixing the limit partition $ \pi $ to its empirical counterpart $ \hat\pi_m $ can induce serious bias.

The computational tractability of the Stephenson-Tawn likelihood was exploited by \cite{Huser2019}, who designed a stochastic expectation-maximisation (EM) algorithm \citep{Nielsen2000, Dempster1977}, in which $ \pi $ is treated as a latent variable and integrated out from the full likelihood by Monte-Carlo integration. More specifically, they sample an ergodic Markov chain of partitions and fit the Stephenson-Tawn likelihood to the data and each partition. The estimates are then averaged over to obtain an approximation of the full likelihood; this method enables likelihood inference in dimensions up to approximately $ D = 20 $. A similar Bayesian approach was proposed by \cite{Thibaud2016}, who also treat $ \pi $ as a latent variable. The authors develop an MCMC algorithm where, in each iteration, they are able to resample partitions conditioned on the data and evaluate the Stephenson-Tawn likelihood. The method was demonstrated on a data set of extreme low temperatures observed at 20 locations.

As an alternative, we propose a \emph{variational inference} \citep{Jordan1999} approach, in which the unknown partition is treated as a latent variable. In contrast to \cite{Huser2019} and \cite{Thibaud2016}, however, we posit a parametric family of partition distributions, from which partitions can be sampled. We then optimise the parameters of the family in conjunction with the Stephenson-Tawn likelihood. Thereby, we find the partition distribution best supported by the data through optimisation, which enables us to perform full likelihood inference without computing the sum over all partitions. 

In a simulation study, we show that our approach does provide accurate parameter estimates in dimensions higher than previous methods in a reasonable amount of time. Furthermore, by using mini-batches of data, our method scales to data sets with a large number of observations without substantially increasing the computational burden. This makes it possible to fit max-stable models to a large number of observations in dimensions up to around 20 to 30 in a reasonable amount of time on a standard desktop computer. The scalability is a major advantage compared to the method of \cite{Huser2019} in which one Markov chain per observation must be sampled. Our method is also well suited for Bayesian analysis, as opposed to the composite likelihood approach, since we obtain a posterior distribution over the partitions. 

The rest of the paper is organised as follows. Section \ref{sec:maxStableProcesses} presents and motivates the use of max-stable models. Section \ref{sec:VI} outlines variational inference and the partition distribution. The results of our simulation study are then presented in Section \ref{sec:experiments} and further discussed in Section \ref{sec:discussion}.

\section{Max-stable processes}\label{sec:maxStableProcesses}
In this section, we briefly describe the theory, models, and inference of max-stable processes used in the study. For a more comprehensive account of max-stable process theory see \citet[Ch. 9]{deHaan2006}, and for an overview of statistical modelling of spatial extremes see \cite{huser2020}.

\subsection{Max-stability}
Max-stable processes extend the univariate generalised extreme value distribution (GEV) to spatial settings. The key property that underpins the use of extreme-value distributions and processes to estimate and extrapolate probabilities of rare events is that of \emph{max-stability}. Let $ X_i(s) $, $ i = 1, 2, \dots $ be independent copies of a random process $ X(s) $ defined on the set of spatial sites $ s \in \mathcal S \subset \mathbb R^2 $. Furthermore, assume that there exist sequences of functions $ a_n(s) > 0 $ and $ b_n(s) $ such that the distributional convergence
\begin{equation}\label{key}
	Z(s) := \lim_{n\rightarrow \infty} \frac{\max\limits_{1\leq i\leq n}\cur{X_i(s)} - b_n(s)}{a_n(s)},
\end{equation}
yields a process $ Z(s) $ that is non-degenerate for all $ s \in \mathcal S $. Then $ Z $ must be max-stable, which means that for each positive integer $ t $ there exist functions $ a_t(s) > 0 $ and $ b_t(s) $, such that if $ \cur{Z_1(s), \ldots, Z_t(s)} $ are i.i.d.\ copies of $ Z(s) $, then
\begin{equation}\label{eq:maxStability}
	\max\cur{Z_1(s), \ldots, Z_t(s)} \overset{d}{=} a_t(s)Z(s)+b_t(s),
\end{equation}
where $ \overset{d}{=} $ denotes equality in distribution. Max-stable processes provide the only possible limits for rescaled point-wise maxima from random processes with non-degenerate margins \citep[Ch. 9]{deHaan2006}.

\subsection{Models}
When constructing models for spatial extreme values it is convenient to express max-stable processes in terms of spectral functions \citep{deHaan1984, Schlather2002}. Let $ P_i $ be points of a Poisson point process on $ (0, \infty) $ with intensity $ r^{-2} \textrm{d}r $, and let $ W_i(s) $ be independent replicates of a non-negative stochastic process with unit mean, indexed by spatial sites $ s \in \mathcal{S} $. Then
\begin{equation}\label{eq:MaxStableRepresentaion}
	Z(s) := \sup_{i} W_i(s)/P_i
\end{equation}
is a max-stable process with unit Fr\'{e}chet margins (i.e.\ $ \Pr(Z(s)<z) = \exp(-1/z),~ z>0 $) and $ D $-dimensional distribution function
\begin{align}
	\Pr\pa{Z(s_1)\leq z_1, \ldots, Z(s_D)\leq z_D} 
	& = \exp\pa{-\E\bra{\max_{k=1,\ldots,D}\pa{\frac{W(s_k)}{z_k}}}} \\
	& =: \exp\bra{-V(z_1,\ldots, z_D)}.\label{eq:MEVDistribution}
\end{align}
The function $ V $ is referred to as the \emph{exponent measure} and summarises the spatial dependence structure. This function is homogeneous of order $ -1 $ (i.e.\ $ V(az_i)=a^{-1}V(z_i) $) and satisfies the marginal constraint $ V(\infty,\ldots,z,\ldots,\infty) = 1/z $ to ensure unit Fr\'{e}chet distributed margins.
A physical interpretation of the spectral function representation, due to \cite{Smith1990a}, is as ``storms'' where $ P_i $ represents the amplitude and $ W_i(s) $ the spatial profile of the storm. 

By specifying the process $ W(s) $ in different ways, a variety of max-stable models can be constructed including the Smith model \citep{Smith1990a}, the Schlather model \citep{Schlather2002}, the Brown-Resnick model \citep{Brown1977, kabluchko2009}, and the extremal-\textit{t} model \citep{Opitz2013}.

In this study we consider two models, starting with the multivariate logistic extreme-value distribution \citep{Gumbel1961}. This is the simplest max-stable distribution, governed by a single parameter $ \theta $ that controls the multivariate dependence. The distribution function is obtained by substituting the exponent measure in \eqref{eq:MEVDistribution} as 
\begin{equation}\label{eq:logisticMaxStable}
	V(z_1,\ldots, z_D) = \pa{\sum_{i=1}^D z_i^{{-1/\theta}}}^\theta,~ 0<\theta\leq 1,
\end{equation}
where $ \theta = 1 $ corresponds to independence and the limiting case $ \theta \rightarrow 0 $ complete dependence. The logistic model is too restrictive for many applications but has an explicit expression for the full likelihood \citep{Shi1995} which can be computed efficiently in high dimensions. Thus, the model serves as a good test case where our estimator can be compared to the maximum likelihood estimator. 

We also consider the Brown-Resnick model which is more flexible than the logistic, and therefore better suited for applications. It is constructed by setting $ W_i(s) = \exp\left (\varepsilon_i(s))-\sigma^2(s) / 2 \right ) $ in \eqref{eq:MaxStableRepresentaion}, where $ \varepsilon_i $ are independent replicates of an intrinsically stationary centred Gaussian process $ \varepsilon $, with $ \varepsilon(0) = 0 $ almost surely. The intrinsic stationarity property ensures that $ \Var\pa{\varepsilon(s) - \varepsilon(s+h)} $ is independent of $ s $, i.e.\ the process $ \varepsilon $ may not be stationary but the increments $ \varepsilon(s) - \varepsilon(s+h) $ are. As the Brown-Resnick model is constructed by a Gaussian process, its density contains multivariate Gaussian distribution functions which need to be approximated. These approximations are computationally burdensome and make the model considerably more demanding to estimate compared to the logistic model.

\subsection{Inference}
Likelihood inference of max-stable processes is complicated by the complex form of the likelihood function. From \eqref{eq:MEVDistribution} the full likelihood for one observation can be obtained as
\begin{equation}\label{eq:fullLikelihood}
	L(\bfz) = \exp\pa{-V(\bfz)}\sum_{\pi\in \mathcal P_D} \prod_{\tau \in \pi}V_{\tau}(\bfz),
\end{equation}
where $ \bfz = (z_1,\ldots,z_D) $. Here, $ \pi = \cur{\tau_1,\ldots,\tau_{\abs{\pi}}} $ denotes a partition of $ \cur{1,\ldots,D} $ and $ \mathcal P_D $ is the set of all partitions, the cardinality of which is the $ D $th Bell number. 
Moreover, $ V_{\tau} $ is the partial derivative of $ V $ with respect to all variables indexed by $ \tau $. Expressions for $ V $ and $ V_{\tau} $ for the Brown-Resnick model can be found in \cite{Huser2013} and \cite{Wadsworth2014}. 

\cite{Stephenson2005} consider $ \pi $ as an observable random variable and therefore instead have the likelihood
\begin{equation}\label{eq:stLikelihood}
	L(\bfz, \pi) = \exp\pa{-V(\bfz)} \prod_{\tau \in \pi}V_{\tau}(\bfz).
\end{equation}
Here, the problematic sum has been reduced to a single term. The authors suggested using the observed partition $ \hat\pi_m $, i.e.\ the partition implied by the occurrence times of the block maxima, as an observation of $ \pi $. \cite{Wadsworth2015}, however, showed that this simplification may induce bias due to model misspecification, especially in scenarios where spatial dependence is weak. Therefore, direct modelling of extreme events with the Stephenson-Tawn likelihood is not preferred. However, by treating $ \pi $ as a latent variable and integrating it out, the relative simplicity of \eqref{eq:stLikelihood} can be used to estimate the full likelihood in \eqref{eq:fullLikelihood}. In the following section, we show how this can be accomplished in a variational inference framework.

\section{Variational inference}\label{sec:VI}
Variational inference is an approximate inference method for estimating latent variable models. The method provides approximate solutions to problems with intractable distributions $ p(\pi\vert x) $, where $ \pi $ are latent and $ x $ observed variables. The idea is to posit a family of parametric distributions $ q $ over $ \pi $, and then, through optimisation, find the distribution that is the closest in \emph{Kullback-Leibler} divergence to the exact conditional distribution $ p(\pi\vert x) $. The distributional family $ q $ is usually referred to as the \emph{variational family} and its parameters the \emph{variational parameters}. Below we describe the method in more detail in relation to the max-stable likelihood in \eqref{eq:fullLikelihood}.

\subsection{Importance weighted auto-encoder estimator}
To make notation easier, we define the full likelihood for the $ i $th observation as
\begin{equation}
	L^i(\bfz;\theta) = \sum_{\pi \in \mathcal P_D} p_\theta^i(\bfz, \pi),
\end{equation}
where $ \pi $ is considered a latent variable. Furthermore, to enhance readability we hereafter exclude the explicit dependence on data, $ \bfz $. By introducing a probability function over the partitions, $ q_\varphi(\pi) $, we may write the likelihood as an expected value,
\begin{equation}
	L^i(\theta) = \sum_{\pi \in \mathcal P_D} \frac{p^i_\theta(\pi)}{q_\varphi(\pi)}q_\varphi(\pi) = \E_{\pi\sim q_\varphi }\bra{\frac{p^i_\theta(\pi)}{q_\varphi(\pi)}} = \E_{\pi_m\overset{iid}{\sim} q_\varphi }\bra{\frac{1}{M}\sum_{m=1}^M\frac{p^i_\theta(\pi_m)}{q_\varphi(\pi_m)}}.
\end{equation}
Here $ M $ specifies the number of partitions that are sampled from $ q_\varphi(\pi) $ when computing the expectation. Since the observations are assumed independent, the log-likelihood of all $ n $ observations is
\[
l(\theta) = \sum_{i=1}^n \log L^i(\theta) = \sum_{i=1}^n \log  \E_{\pi_m\overset{iid}{\sim} q_\varphi }\bra{\frac{1}{M}\sum_{m=1}^M\frac{p^i_\theta(\pi_m)}{q_\varphi(\pi_m)}}.
\]
Using Jensen's inequality, we find that
\begin{equation}\label{eq:iwae}
	l^i(\theta):= \log L^i(\theta) \geq \E_{\pi_m\overset{iid}{\sim} q_\varphi }\bra{\log \frac{1}{M}\sum_{m=1}^M\frac{p^i_\theta(\pi_m)}{q_\varphi(\pi_m)}}=:\mathcal L_M^i(\theta,\varphi).
\end{equation}
The right-hand side is known as the \emph{importance weighted auto-encoder estimator} (IWAE) \citep{Burda2016}, and in the special case $ M=1 $ this is the \emph{evidence lower bound} (ELBO) \citep{Jordan1999}, which provides a lower bound for the log-likelihood. We get the bound for the complete sample as
\begin{align}
	l(\theta) = \sum_{i=1}^{n} l^i(\theta) \geq \sum_{i=1}^{n} \mathcal L_M^i(\theta,\varphi) 
	& = \sum_{i=1}^{n} \E_{\pi_m\overset{iid}{\sim} q_\varphi }\bra{\log \frac{1}{M}\sum_{m=1}^M\frac{p^i_\theta(\pi_m)}{q_\varphi(\pi_m)}} \\
	& = \E_{\pi_m\overset{iid}{\sim} q_\varphi }\bra{\sum_{i=1}^{n}\log \frac{1}{M}\sum_{m=1}^M\frac{p^i_\theta(\pi_m)}{q_\varphi(\pi_m)}}.\label{eq:boundFullSample}
\end{align}
\cite{Burda2016} showed that this bound can be made arbitrarily tight, i.e.\ brought closer to the log-likelihood, by increasing $ M $. Increasing $ M $, however, also increases the computational burden. Furthermore, \cite{Rainforth2018} showed that while larger $ M $ indeed yields a tighter bound, it also increases the variance of the gradient estimate with respect to $\varphi$, which can make the optimisation more difficult. The authors, however, suggest that there may be a ``sweet spot'' for $ M $ that balances the tightness of the bound and the variance of the gradient estimates. As such, one must determine a suitable value of $ M $ to obtain accurate parameter estimates. This is discussed further in relation to the max-stable models in Section \ref{sec:implementation}.

To estimate the parameters of the max-stable and variational models we want to maximise the expected value \eqref{eq:boundFullSample} with respect to $\theta$ and $\varphi$. This will be accomplished using stochastic gradient ascent and hence, we calculate the gradients with respect to the parameters:
\begin{align}
	\partial_\theta \mathcal L_M^i(\theta,\varphi)
	&=  \E_{\pi_m\overset{iid}{\sim} q_\varphi }\bra{\partial_\theta \log  \sum_{m=1}^M\frac{p^i_\theta(\pi_m)}{q_\varphi(\pi_m)}}\label{eq:firstGrad}\\
	&=  \E_{\pi_m\overset{iid}{\sim} q_\varphi }\bra{\pa{ \sum_{m=1}^M\frac{p^i_\theta(\pi_m)}{q_\varphi(\pi_m)}}^{-1}  \sum_{m=1}^M\frac{\partial_\theta p^i_\theta(\pi_m)}{q_\varphi(\pi_m)} }\\
	&=  \E_{\pi_m\overset{iid}{\sim} q_\varphi }\bra{\pa{ \sum_{m=1}^M\frac{p^i_\theta(\pi_m)}{q_\varphi(\pi_m)}}^{-1}  \sum_{m=1}^M\frac{p^i_\theta(\pi_m)}{q_\varphi(\pi_m)} \partial_\theta \log p^i_\theta(\pi_m)}.
\end{align}
Then it is clear that if we sample $\pi_m \sim q_\varphi$, the quantity inside the expectation will give an unbiased estimate of $\partial_\theta \mathcal L_M^i$. 

Further,
\begin{align}
	\partial_\varphi \mathcal L_M^i(\theta,\varphi)
	= & \partial_\varphi \sum_{\pi_1,\ldots ,\pi_M\in \mathcal P_D} \log \sum_{m=1}^M\frac{p^i_\theta(\pi_m)}{q_\varphi(\pi_m)} q_\varphi(\pi_1)\cdots q_\varphi(\pi_M) \\
	= &  \sum_{\pi_1,\ldots ,\pi_M\in \mathcal P_D} q_\varphi(\pi_1)\cdots q_\varphi(\pi_M) \partial_\varphi \log \sum_{m=1}^M\frac{p^i_\theta(\pi_m)}{q_\varphi(\pi_m)}  \\
	& + \sum_{\pi_1,\ldots ,\pi_M\in \mathcal P_D}  \log \sum_{m=1}^M\frac{p^i_\theta(\pi_m)}{q_\varphi(\pi_m)} \partial_\varphi q_\varphi(\pi_1)\cdots q_\varphi(\pi_M) \\
	= & \E_{\pi_m\overset{iid}{\sim} q_\varphi }\bra{\partial_\varphi \log \sum_{m=1}^M\frac{p^i_\theta(\pi_m)}{q_\varphi(\pi_m)}} \\
	& +  \E_{\pi_m\overset{iid}{\sim} q_\varphi }\bra{ \log\pa{ \sum_{m=1}^M\frac{p^i_\theta(\pi_m)}{q_\varphi(\pi_m)}} \sum_{m=1}^M \partial_\varphi \log q_\varphi(\pi_m)} \\
	=& -\E_{\pi_m\overset{iid}{\sim} q_\varphi }\bra{\pa{ \sum_{m=1}^M\frac{p^i_\theta(\pi_m)}{q_\varphi(\pi_m)}}^{-1}\sum_{m=1}^M\pa{\frac{p^i_\theta(\pi_m)}{q_\varphi(\pi_m)}\partial_\varphi\log q_\varphi(\pi_m)}} \\
	&+  \E_{\pi_m\overset{iid}{\sim} q_\varphi }\bra{ \log\pa{ \sum_{m=1}^M\frac{p^i_\theta(\pi_m)}{q_\varphi(\pi_m)} }\sum_{m=1}^M \partial_\varphi \log q_\varphi(\pi_m)}.\label{eq:lastGrad}
\end{align}
Again, this allows us to generate unbiased samples from $\partial_\varphi \mathcal L_M^i$.

\subsection{Variational family}

Next, we need to choose a variational family of distributions for the partitions, $ q_\varphi(\pi) $. This family should be rich enough to contain a partition distribution that makes the inequality in \eqref{eq:iwae} tight, while still being sufficiently simple to enable efficient optimisation. With this in mind, we use the \emph{Evans-Pitman attraction} (EPA) distribution \citep{Dahl2017}, which provides a way to construct partitions sequentially while accounting for pairwise similarities between the items to be partitioned. More specifically, define a partition as $ \pi = \cur{\tau_1,\ldots,\tau_{\lvert\pi\rvert}} $ where $ \tau_i $ are the subsets that constitute the partition and $ \lvert\pi\rvert $ is the number of subsets. We want to partition the items of the set $ \cur{I_1,\ldots,I_D} $ into subsets $ \cur{\tau_1,\ldots,\tau_{\lvert\pi\rvert}} $. This is accomplished as follows: in step $ t = 1 $, item $ I_1 $ is assigned to subset $ \tau_1 $ with probability $ 1 $. At step $ t = 2 $, item $ I_2 $ is assigned with certain probabilities to either $ \tau_1 $ or $ \tau_2 $. More generally, at step $ t $, item $ I_t $ is assigned to either one of the $ \lvert\pi_{t-1}\rvert $ existing subsets in the partition $ \pi_{t-1} $, or to a new subset according to the probabilities
\begin{align}
	q_t(\alpha, \delta, \lambda, \pi_{t-1}) &= \Pr(I_t \in \tau \mid \alpha, \delta, \lambda, \pi_{t-1}) \\
	&= 
	\begin{cases}
		\frac{t-1-\delta \lvert\pi_{t-1}\rvert}{\alpha + t-1} \cdot \frac{\sum_{I_s \in \tau} \lambda(I_t, I_s)}{\sum_{s=1}^{t-1}\lambda(I_t, I_s)} & \textrm{for $ \tau $} \in \pi_{t-1} \\
		\frac{\alpha + \delta \lvert \pi_{t-1} \rvert}{\alpha+t-1} & \textrm{for $ \tau $ being a new subset}.
	\end{cases}
\end{align}
The parameters $ \delta \in [0, 1) $ and $ \alpha > -\delta $ control the number of subsets in the partition. Furthermore, $ \lambda $ is a similarity function defined as $ \lambda(i,j) = f(d_{ij}) $, for some non-increasing function $ f $ of pairwise distances $ d_{ij} $ between items $ I_i$ and $ I_j $. We use the exponential similarity $ \lambda_\rho(i,j) = \exp(-d_{ij}/\rho) $ with parameter $ \rho > 0 $, and define the distance as 
\begin{equation}\label{eq:distance}
	d_{ij} = \abs{z(s_i)-z(s_j)},
\end{equation}
i.e.\ the absolute difference between observation values from sites $ s_i $ and $ s_j $, $ i,j = 1, \dots, D $. This definition of $ d_{ij} $ rests on the assumption that simultaneous extreme values observed at different sites are likely to result from the same extremal event. It should be noted that, while the parameters $ \alpha $ and $ \delta $ are shared between all observations, a consequence of using the distance definition in \eqref{eq:distance} is that the distance matrix will be different for each observation. An alternative definition of $ d_{ij} $ is the Euclidean distance between sites, which implicitly assumes that the closer two sites are, the more likely it is to observe simultaneous extreme values at them. With this latter definition, the distance matrix is shared between all observations because the sites are fixed. Which is the best choice of distance with respect to tightening \eqref{eq:iwae} is however an empirical question that we choose to not investigate further.

The EPA distribution was chosen because it has a closed-form expression for the probability mass function and is easy to sample partitions from. Furthermore, the similarity function $ \lambda $ provides a direct way to incorporate information on pairwise similarities between items in the allocation process. This is useful in an analysis of spatial extreme values where a single extreme event might yield maxima at multiple sites. Related partition distributions are the Chinese restaurant process (CRP) of \cite{Aldous1985} and the distance-dependent Chinese restaurant process (ddCRP) of \cite{Blei2011}. The CRP, however, fails to incorporate information on pairwise distances, and while the ddCRP utilises pairwise distance information, it is less flexible than the EPA distribution because it does not have a discount parameter $ \delta $.

\section{Numerical experiments}\label{sec:experiments}
The performance of our variational inference estimator $ \hat \theta_\textrm{VI} $, of the vector of max-stable model parameters $ \theta $, is investigated in a simulation study where the statistical properties and computational efficiency are assessed using first, the logistic model, and second, the Brown-Resnick model. With the logistic model we can compare $ \hat \theta_\textrm{VI} $ to the maximum likelihood estimator $ \hat \theta_\textrm{MLE} $ in high dimensions; the results are presented in Section \ref{sec:logistic}. In Section \ref{sec:brown-resnick}, results from the Brown-Resnick model are presented, which illustrates how our estimator performs with a more complex model that is better suited for applications.

\subsection{Implementation}\label{sec:implementation}
Here we describe the general implementation of the numerical experiments. The model-specific details will be presented in their respective sections. The simulations are carried out in Julia 1.7 \citep{Julia2017} and all computations are performed on standard desktop computers. The code is available at \url{https://github.com/LPAndersson/MaxStableVI.jl}. 

The computational times reported below represent experiments computed on a single CPU core at clock frequency $ 3.40 $ GHz. We use stochastic gradient ascent (SGA) with momentum to estimate the max-stable model parameters, and standard SGA to estimate the variational parameters. The derivatives inside the expected values in \eqref{eq:firstGrad} - \eqref{eq:lastGrad} are evaluated using automatic differentiation. 

\begin{figure}[tp]
	\centering
	\includegraphics[width=0.75\linewidth]{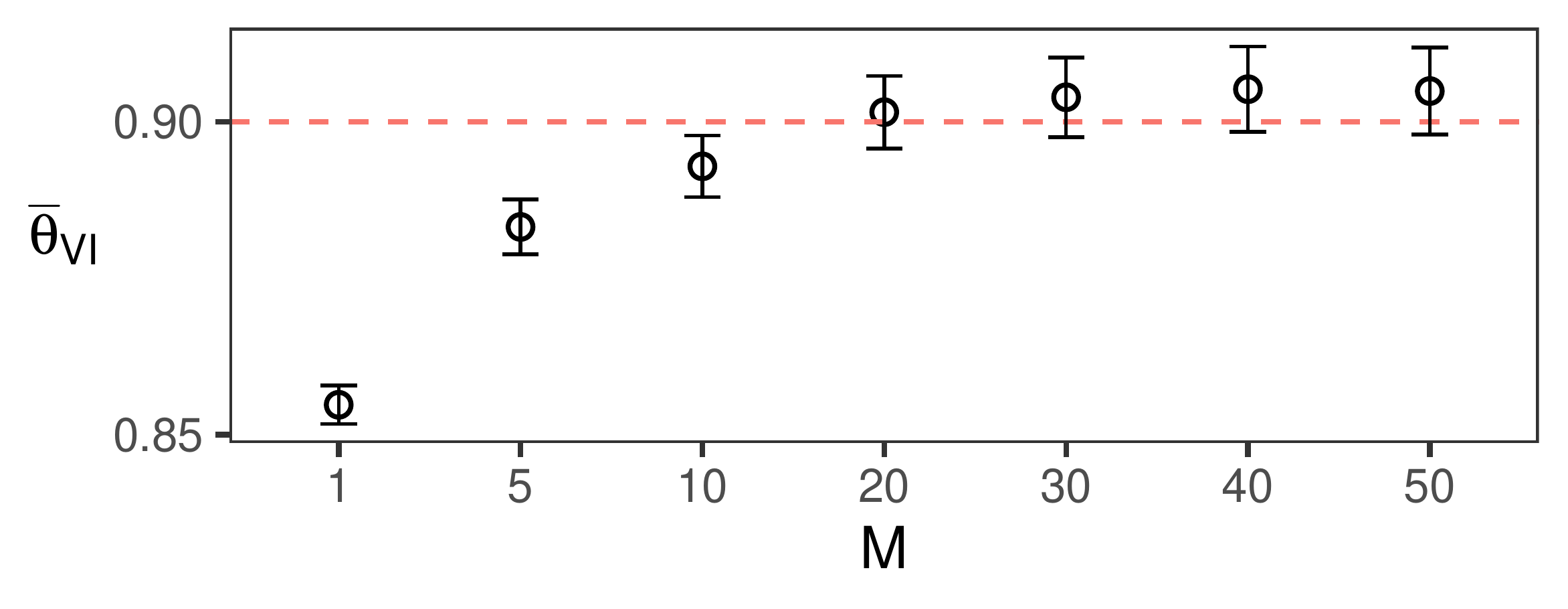} 
	\caption{Averages of $ \hat \theta_\textrm{VI} $ with associated $ 95\% $ confidence intervals for different values of $ M $, computed from $ 100 $ replications based on the logistic model with $ \theta = 0.9 $, $ D = 10 $ and $ n=20 $. The optimiser was run for $ 5000 $ iterations and the initial value of $ \theta $ was set to $ 0.6 $. As $ q_\varphi(\pi) $ we used the Evans-Pitman attraction distribution with similarity function $ \lambda_\rho(i,j) = \exp(-\abs{z(s_i)-z(s_j)}/\rho) $. The results are similar when $ \theta = 0.3 $} \label{fig:logistic_m}
\end{figure}

The suitable number of samples $ M $ in the IWAE estimator is assessed through simulation to balance estimation accuracy and computational time. To illustrate how the choice of $ M $ may affect the estimation accuracy, we draw observations distributed according to the logistic model in dimension $ D = 10 $ with $ \theta = 0.9 $ and $ n = 20 $ temporal replicates, and compute $ \hat \theta_\textrm{VI} $ for each $ M \in \{1, 5, 10, 20, 30, 40, 50\} $. This is repeated for $ 100 $ replications and the average estimates with associated $ 95\% $ confidence intervals are presented in Figure \ref{fig:logistic_m}. There is a clear trade-off between bias and computation time, where, in the presented scenario, increasing $ M $ reduces bias up to around $ M = 20 $. Thereafter further increases only raise the computational burden.

\subsection{Logistic model results}\label{sec:logistic}
\begin{figure}[tp]
\includegraphics[width=1\linewidth]{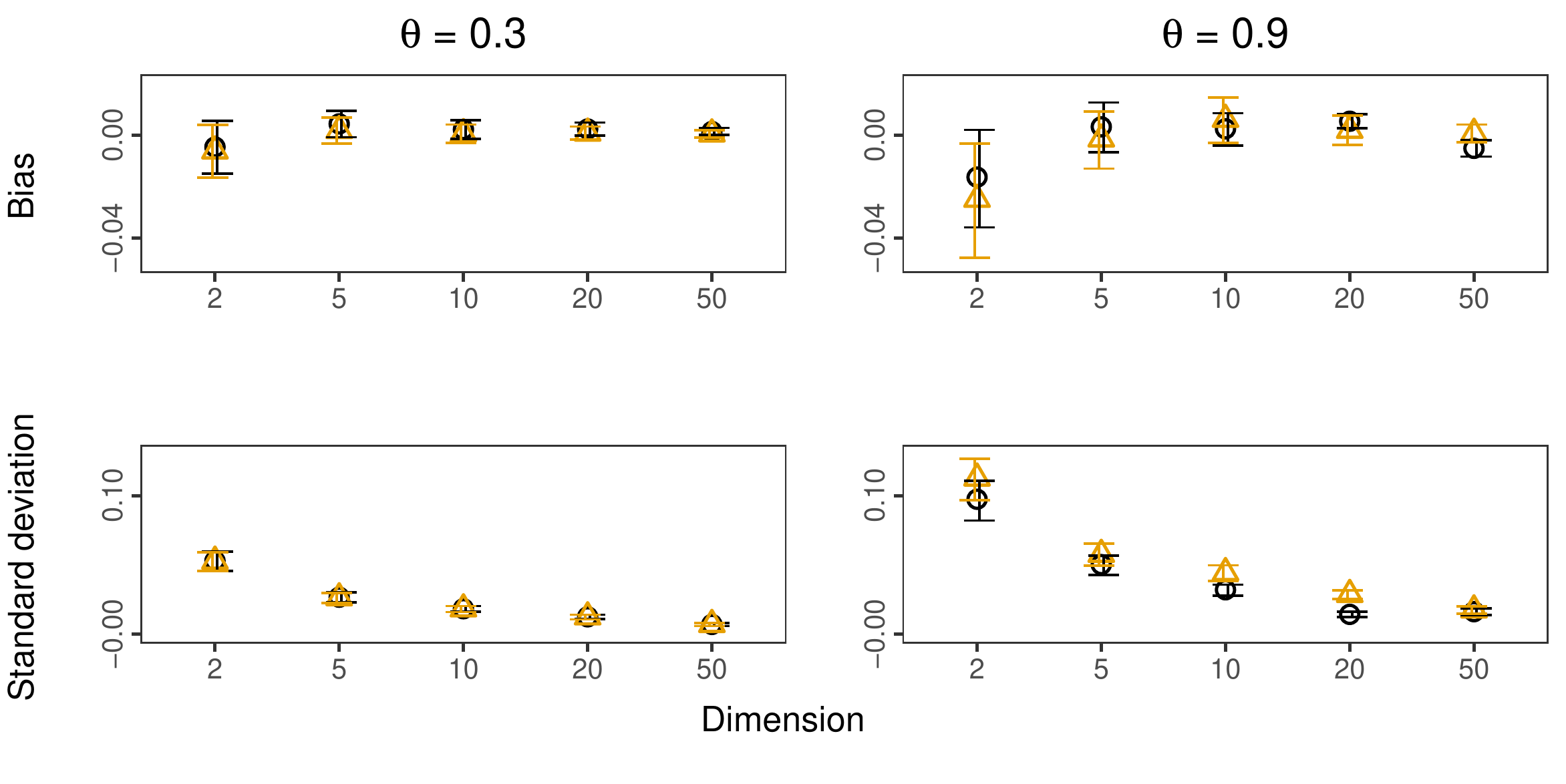} 
\caption{Estimated bias and standard deviation of $ \hat \theta_\textrm{VI} $ (circles, black) and $ \hat \theta_\textrm{MLE} $ (triangles, orange) with $ 95\% $ bootstrap percentile confidence intervals, computed from $ 100 $ replications based on the logistic model with $ n = 20 $ and $ M = 25 $. The optimiser was run for $ 5000 $ iterations and the initial value of $ \theta $ was set to $ 0.6 $. As $ q_\varphi(\pi) $ we used the Evans-Pitman attraction distribution with similarity function $ \lambda_\rho(i,j) = \exp(-\abs{z(s_i)-z(s_j)}/\rho) $} \label{fig:logistic_statProp}
\end{figure}

To assess the properties of $ \hat \theta_\textrm{VI} $, we generate logistic random vectors in dimensions $ D \in \{2, 5, 10, 20, 50\} $ with $ n = 20 $ temporal replicates, for $ \theta \in \{0.3, 0.9\} $ (strong, weak dependence), resulting in $ 10 $ scenarios. These scenarios together give a good overview of how our estimator performs in different dimensions under varying dependence strengths. The optimiser is run for $ R = 5000 $ iterations to ensure that all replicates show convergence, and the learning rate of the SGA is tuned separately for model and guide parameters in each scenario. The number of partitions sampled from $ q_\varphi(\pi) $ in each iteration is tuned to $ M = 25 $, and the starting value of $ \theta $ is set to $ 0.6 $. To estimate the distribution of $ \hat \theta_\textrm{VI} $ we simulate $ 100 $ replications. The results are presented in Figure \ref{fig:logistic_statProp} with corresponding results for $ \hat \theta_\textrm{MLE} $. We see that the bias of $ \hat \theta_\textrm{VI} $ is generally low and close to that of $ \hat \theta_\textrm{MLE} $, both under strong and weak dependence. The standard deviation is higher under weak dependence but decreases with increasing dimension under both dependence strengths.

\begin{figure}[tp]
	\includegraphics[width=1\linewidth]{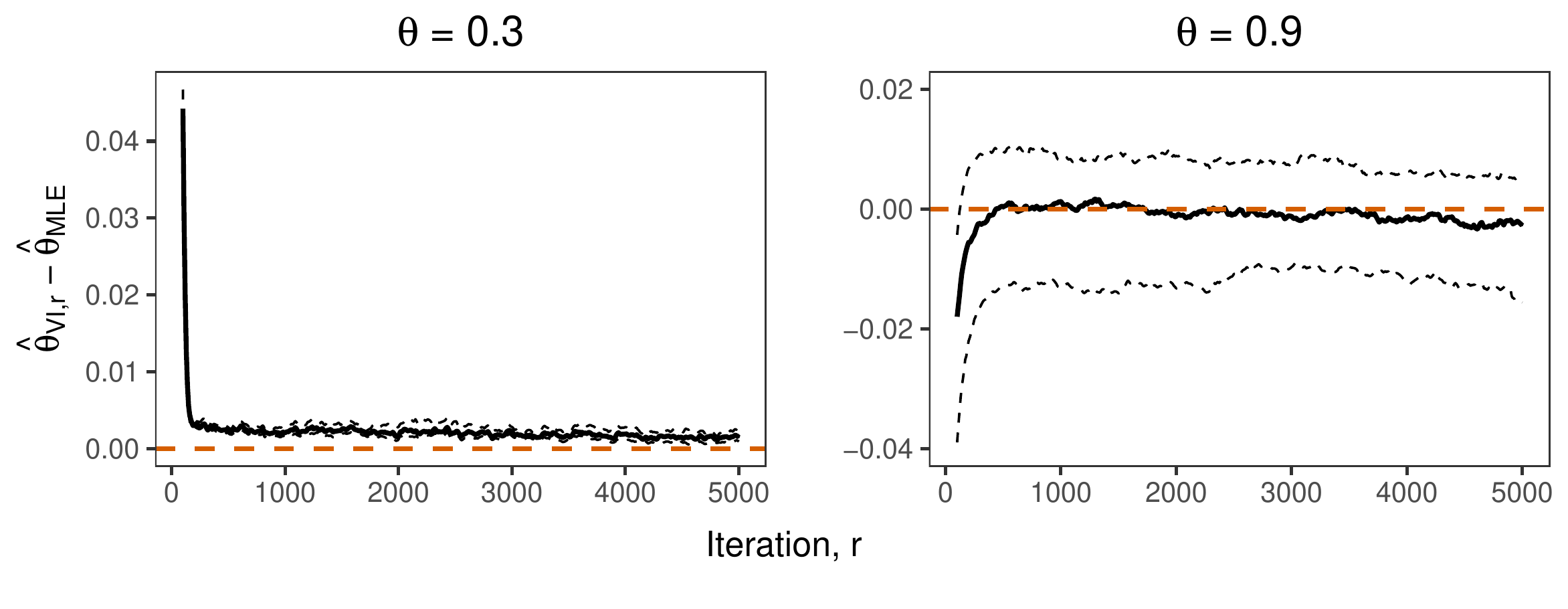} 
	\caption{Median (solid line) with $ 25 $th and $ 75 $th quantiles (dashed lines) of centred parameter traces $ \hat \theta_\textrm{VI, r} - \hat \theta_\textrm{MLE} $ as functions of optimisation iterations $ r = 1, \dots, 5000 $ computed from $ 100 $ independent replications based on the logistic model with $ D = 10 $, $ n = 20 $ and $ M = 25 $. As $ q_\varphi(\pi) $ we used the Evans-Pitman attraction distribution with similarity function $ \lambda_\rho(i,j) = \exp(-\abs{z(s_i)-z(s_j)}/\rho) $. The first $ 10 $ iterations were excluded to enhance readability} \label{fig:logistic_quantMle}
\end{figure}

Next, we consider the computational efficiency. To investigate how fast the optimisation converges we compute traces of the centred parameter $ \hat \theta_\textrm{VI,r} - \hat \theta_\textrm{MLE} $, $ r = 1, 2, \dots, R $. Figure \ref{fig:logistic_quantMle} shows the quartiles of these traces as a function of the optimisation iterations $ r $ for the scenarios with $ \theta \in \{0.3, 0.9\} $ and $ D = 10 $. The results for other dimensions are similar. The median traces approach $ 0 $ fast and one may use fewer iterations than $ 5000 $ while still obtaining accurate parameter estimates. Concerning actual computational time, this depends on, amongst other things, the choice of $ M $, learning rates, and hardware. But to give a rough measure, computing one estimate $ \hat \theta_\textrm{VI} $ with $ \theta = 0.9 $, $ R = 1000 $ with the remaining settings unchanged takes about $ 15.2 $ minutes for $ D = 20 $, $ 44.5 $ minutes for $ D = 50 $ and $ 2.3 $ hours for $ D = 100 $. 

The computational time is roughly linear in $ M $, although, while larger $ M $ increases computational burden, it also enables a higher learning rate and fewer iterations until convergence. Our experience from the simulation study indicates that a rather large $ M $ with a high learning rate and few iterations yields the fastest convergence. Further reduction of the computational time can also be achieved by parallelising the computations for the $ M $ sampled partitions. 

To summarise, the results show that our estimator yields accurate parameter estimates in a reasonable amount of time in dimension $ D = 100 $ with the settings of the simulation study. The computational time can be reduced further by potentially using fewer optimisation iterations, and also by running the computations for the $ M $ sampled partitions in parallel.

\subsection{Brown-Resnick model results}\label{sec:brown-resnick}
We now turn our attention to the more complex Brown-Resnick model. The intrinsically stationary Gaussian processes that define this model are characterised by their semivariogram $ \gamma(h) = \frac{1}{2}~\E\bra{\pa{\varepsilon(s) - \varepsilon(s+h)}^2} $. Here, we use the commonly used isotropic semivariogram $ \gamma(h) = \gamma(s_1,s_2) = \pa{ \lVert s_1 - s_2 \rVert / \lambda }^\nu $ where $ \lVert \cdot \rVert $ is the Euclidean norm, and $ \lambda > 0 $ and $ \nu \in (0,2] $ constitute range and smoothness parameters. Thus, the model has two parameters, and, furthermore, the spatial dependence is completely determined by the distances between the sites $ s $. As the density function contains Gaussian distribution functions of dimension up to $ D-1 $ the Brown-Resnick model is much more computationally demanding than the logistic model. 

To assess the performance of our estimator $ (\hat \lambda_\textrm{VI}, \hat \nu_\textrm{VI}) $, we generate $ 100 $ independent vectors randomly at $ D = 5 $ sites in $ [0, 1]^2 $ with $ n = 10 $ temporal replicates. Parameter values are set to $ \lambda \in \{0.5, 1.5\} $ (weak, strong dependence) and $ \nu \in \{0.5, 1.5\} $ (rough, smooth process), which results in the four scenarios presented in Table \ref{tab:scenarios}. The rather low dimension and few temporal replicates are chosen such that $ (\hat \lambda_\textrm{MLE}, \hat \nu_\textrm{MLE}) $ can be computed in a reasonable amount of time and used as a baseline for our estimates. Since our goal is to perform full likelihood inference we want our estimates to resemble those of the MLE. The optimiser is run for $ R = 2000 $ iterations with learning rates tuned for each scenario. The number of partitions sampled from $ q_\varphi(\pi) $ is tuned to $ M = 50 $, and the starting values for both $ \lambda $ and $ \nu $ are set to $ 1 $. 
\begin{table}
	\centering
	\caption{Scenarios for the Brown-Resnick model with parameters $ \lambda, \nu $.}
	\label{tab:scenarios}
	\begin{tabular}{ccccc}
		\hline
		Scenario & 1 & 2 & 3 & 4 \\
		\hline
		$ \lambda	$ & $ 0.5 $ & $ 0.5 $ & $ 1.5 $ & $ 1.5 $ \\
		$ \nu		$ & $ 0.5 $ & $ 1.5 $ & $ 0.5 $ & $ 1.5 $ \\
		\hline
	\end{tabular}
\end{table}

\begin{figure}[tp]
	\includegraphics[width=1\linewidth]{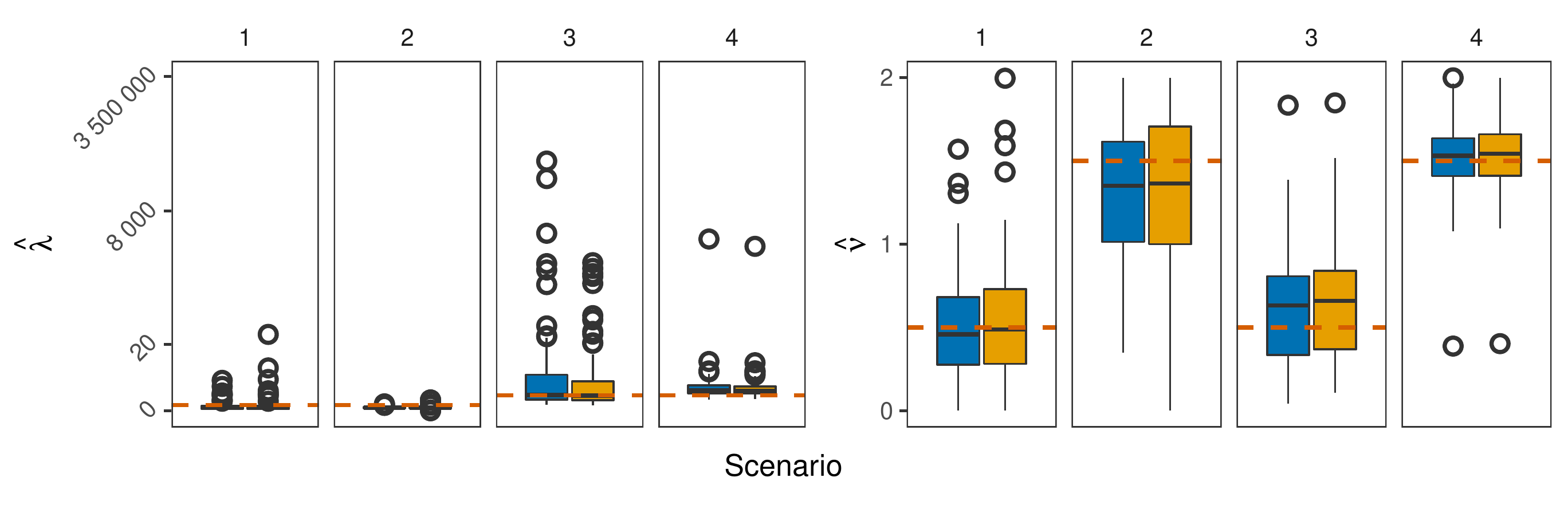} 
	\caption{Estimates from $ (\hat \lambda_\textrm{VI}, \hat \nu_\textrm{VI}) $ (left boxes, blue) and $ (\hat \lambda_\textrm{MLE}, \hat \nu_\textrm{MLE}) $ (right boxes, yellow) for each scenario in Table \ref{tab:scenarios}, from $ 100 $ simulations based on the Brown-Resnick model with semivariogram $ \gamma(h) = \pa{ \lVert h \rVert / \lambda }^\nu $ simulated at $ D = 5 $ sites in $ [0, 1]^2 $ with $ n = 10 $, $ M = 50 $ and $ R = 2000 $ optimiser iterations. Initial values of $ \lambda $ and $ \nu $ were set to $ 1 $. As $ q_\varphi(\pi) $ we used the Evans-Pitman attraction distribution with similarity function $ \lambda_\rho(i,j) = \exp(-\abs{z(s_i)-z(s_j)}/\rho) $. Dashed orange lines show true parameter values. From scenarios $ 1 $ and $ 3 $ we excluded $ 10 $ respectively $ 5 $ values with very large $ \hat \lambda_\textrm{MLE} $ to enhance readability} \label{fig:br_boxplots}
\end{figure}

The distributions of parameter estimates are presented as boxplots in Figure \ref{fig:br_boxplots}. Overall, the simulated distributions of $ \hat \lambda_\textrm{VI} $ and $ \hat \nu_\textrm{VI} $ compare well to those of $ \hat \lambda_\textrm{MLE} $ and $ \hat \nu_\textrm{MLE} $, and the bias is generally low. The estimation variability is quite high due to the small data sets, and from scenarios 1 and 3 we excluded 5 and 10 replicates, respectively, with very large $ \hat \lambda_\textrm{MLE} $ to enhance readability. In addition to the high variability, another complicating factor is that the estimated parameters affect the dependence strength in opposing directions, and, are in that sense negatively correlated. Hence, small values of $ \hat \nu $ coincide with large values of $ \hat \lambda $, and also, small changes in $ \hat \nu $ yield large changes in the value of $ \hat \lambda $ for which the likelihood is maximised. From scenarios 1, 2, and 3 we needed to run 9, 9, and 3 replicates, respectively, for more than 2000 iterations to see convergence. Moreover, for 6, 4, and 5 replicates from scenarios 1, 2, and 3 our estimator failed to converge. The reason for this non-convergence is not entirely clear, but we hypothesize that it is an effect of the complications described above.

\begin{figure}[tp]
	\includegraphics[width=1\linewidth]{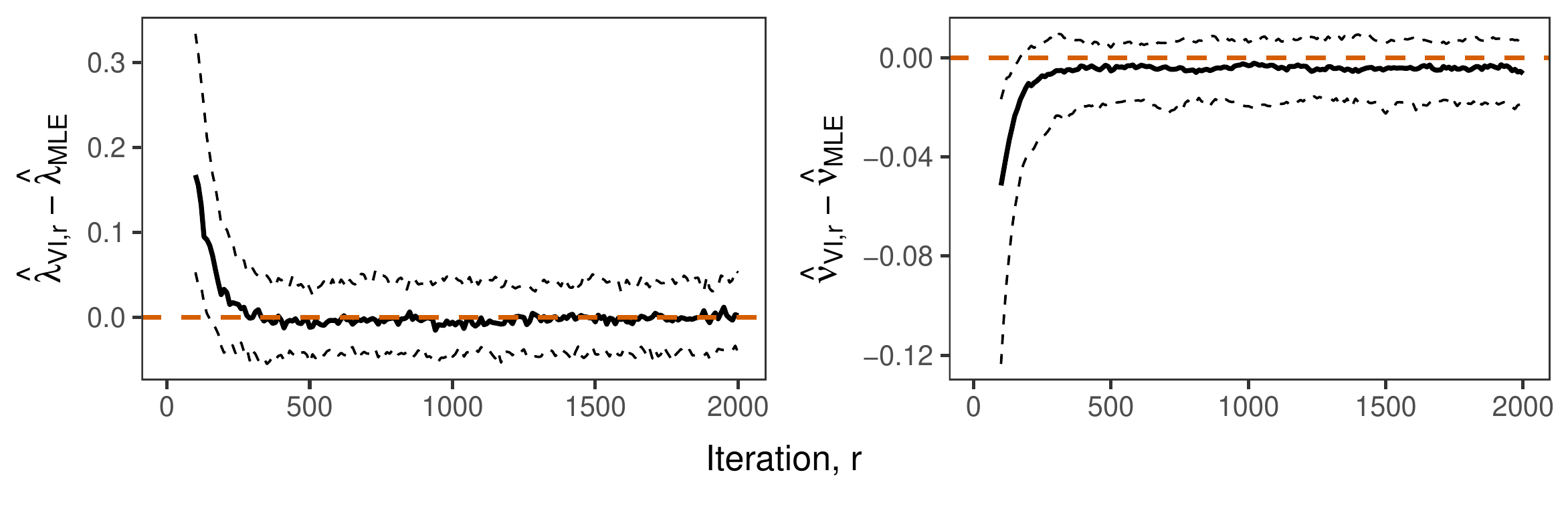} 
	\caption{Median (solid line) with $ 25 $th and $ 75 $th quantiles (dashed lines) of the centred parameter traces $ \hat \lambda_\textrm{VI, r} - \hat \lambda_\textrm{MLE} $ and $ \hat \nu_\textrm{VI, r} - \hat \nu_\textrm{MLE} $ as functions of the optimiser iterations $ r = 1, \dots, 2000 $, from $ 100 $ simulations based on the Brown-Resnick model with semivariogram $ \gamma(h) = \pa{ \lVert h \rVert / \lambda }^\nu $ simulated at $ D = 5 $ sites in $ [0, 1]^2 $ with $ \lambda = \nu = 1.5 $, $ D = 5 $, $ n = 10 $ and $ M = 50 $. Initial values of $ \lambda $ and $ \nu $ were set to $ 1 $. As $ q_\varphi(\pi) $ we used the Evans-Pitman attraction distribution with similarity function $ \lambda_\rho(i,j) = \exp(-\abs{z(s_i)-z(s_j)}/\rho) $. The first $ 10 $ iterations were excluded to enhance readability} \label{fig:br_quartTraceN10}
\end{figure}

We will now examine the speed of convergence. Figure \ref{fig:br_quartTraceN10} shows the quartiles of centred parameter traces $ \hat \lambda_\textrm{VI, r} - \hat \lambda_\textrm{MLE} $ and $ \hat \nu_\textrm{VI, r} - \hat \nu_\textrm{MLE} $, $ r = 1,2,\dots,R $, with the same settings as described previously. True parameter values are $ \lambda = \nu = 1.5 $. The medians of the estimates approach $ 0 $ quickly, suggesting that fewer optimisation iterations may be used while retaining sufficient accuracy of the parameter estimates. We also calculate the time it takes to compute one estimate in the scenario $ \lambda = \nu = 1.5 $ with $ R = 1000 $ optimisation iterations and the remaining settings unchanged. This takes about $ 3 $ hours for $ D = 5 $, $ 9.9 $ hours for $ D = 10 $, $ 48.8 $ hours for $ D = 20 $,  and $ 142.5 $ hours for $ D = 30 $. The main bottleneck is the computation of multivariate Gaussian distribution functions, which is here carried out with a quasi-Monte Carlo algorithm by \cite{Genz1992}. Faster computation of these Gaussian probabilities would greatly reduce the computational time of our estimator.

\begin{figure}[tp]
	\includegraphics[width=1\linewidth]{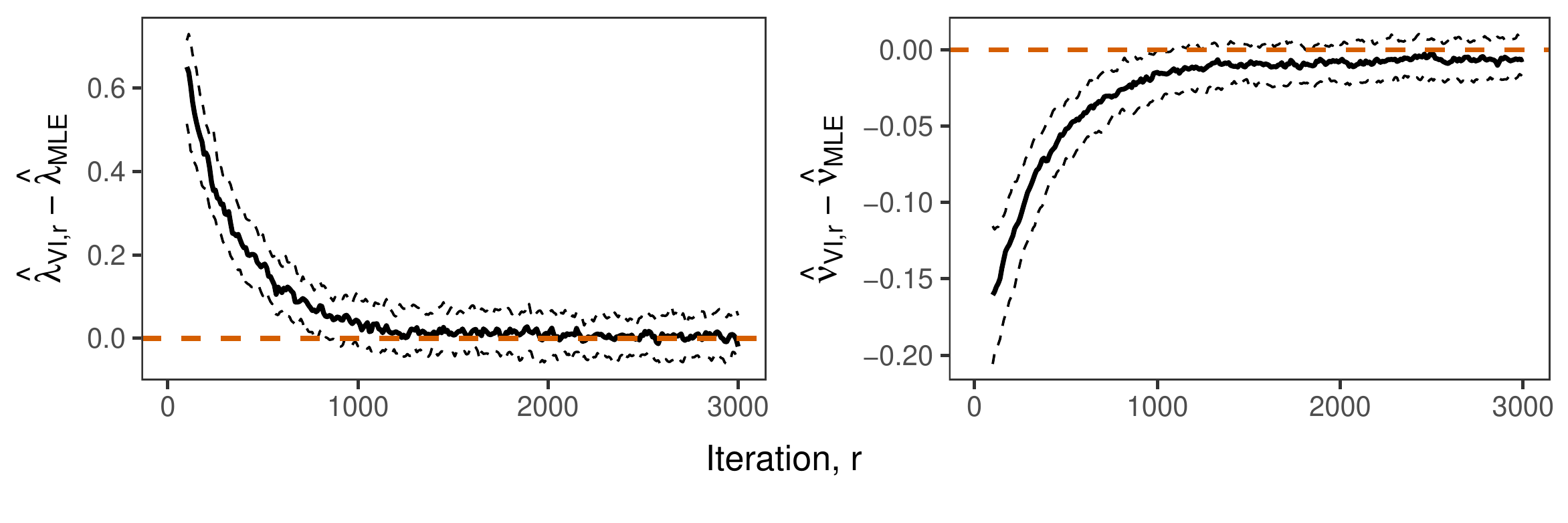} 
	\caption{Median (solid line) with $ 25 $th and $ 75 $th quantiles (dashed lines) of the centred parameter traces $ \hat \lambda_\textrm{VI, r} - \hat \lambda_\textrm{MLE} $ and $ \hat \nu_\textrm{VI, r} - \hat \nu_\textrm{MLE} $ as functions of the optimiser iterations $ r = 1, \dots, 3000 $, from $ 100 $ simulations based on the Brown-Resnick model with semivariogram $ \gamma(h) = \pa{ \lVert h \rVert / \lambda }^\nu $ simulated at $ D = 5 $ sites in $ [0, 1]^2 $ with $ \lambda = \nu = 1.5 $, $ D = 5 $, $ M = 50 $ and $ n = 150 $ with mini-batches of size $ 10 $. Initial values of $ \lambda $ and $ \nu $ were set to $ 1 $. As $ q_\varphi(\pi) $ we used the Evans-Pitman attraction distribution with similarity function $ \lambda_\rho(i,j) = \exp(-\abs{z(s_i)-z(s_j)}/\rho) $. The first $ 10 $ iterations were excluded to enhance readability} \label{fig:br_quartTraceN150}
\end{figure}

One major advantage of our method is that it can be readily applied to data sets with a large number of temporal replicates, because the optimisation updates can be computed with mini-batches of data. This feature makes our method directly useful in applications, where data sets with $ 50 $ to $ 100 $ observations are common. Figure \ref{fig:br_quartTraceN150} shows the quartiles of centred parameter traces for the same settings described previously, but with $ n = 150 $ and where the optimisation iterations are computed with random mini-batches of size $ 10 $. The number of optimiser iterations is set to $ R = 3000 $. Convergence is slower than when $ n = 10 $, but the computational time is considerably shorter than when all $ 150 $ temporal replicates are used in each iteration.

To summarise, the results from the simulation study suggest that our estimator provides accurate estimates of the Brown-Resnick model parameters in dimension $ D = 30 $ in a reasonable amount of time. This is an improvement compared to the method of \cite{Huser2019}, where the stochastic EM algorithm enables inference in dimensions up to around $ D = 20 $. Furthermore, our method has the advantage of scaling to data sets with a large number of observations without substantially increasing the computation time, as the optimisation updates can be computed using mini-batches of data. Altogether, with our method the Brown-Resnick model can be fitted to data sets of dimensions up to 20-30 with a large number of observations on a standard desktop computer in a reasonable amount of time. Still, some convergence issues were encountered in the simulation study and further improvements to the method could be made to alleviate these issues. 

\section{Discussion}\label{sec:discussion}
We propose a variational inference estimator for full likelihood inference of max-stable processes that circumvents the need to compute the sum over all partitions of data. This method also avoids potential model misspecification issues caused by fixing the partition, by instead treating the unknown partition as a latent variable and positing a parametric family of partition distributions. The parameters of the partition distribution family are then optimised in conjunction with the joint max-stable likelihood of the partition and data. In a simulation study, we show that our estimator provides accurate parameter estimates of the logistic model in dimension $ D = 100 $ in 2 to 3 hours. We can also fit the Brown-Resnick model in dimensions up to around $ D = 30 $ in a reasonable amount of time. Furthermore, by using mini-batches of data, our method can be applied to data sets with a large number of observations without substantially increasing the computational time. This enables us to fit max-stable models to data sets of dimensions up to 20-30 with a large number of observations on a standard desktop computer in a reasonable amount of time. The scalability to more observations is a major advantage of our method in comparison to previous methods. Moreover, our method can be applied to any max-stable model with known expressions for $ V $ and its partial derivatives. 

Some convergence issues and very large estimates were observed in the simulation study and, while part of these issues seem to be a result of high estimation variability due to small data sets, further improvements to the method would be needed to alleviate other issues. Possible improvements include finding an orthogonal parametrisation of the Brown-Resnick model or positing a different partition distribution. Regarding computation time, further decreases can be achieved by performing the computations for the $ M $ sampled partitions in parallel, and potentially by defining a stopping criteria for the optimiser. Another question for future research is how to systematically determine a suitable value for $ M $.

\bibliographystyle{apalike}
\bibliography{bibliography}

\end{document}